\documentclass[a4paper,11pt]{article}
\pdfoutput=1 
\usepackage{jcappub} 
\usepackage[T1]{fontenc} 
\usepackage{multirow} 
\usepackage{multicol} 
\usepackage{enumitem} 
\usepackage{array}
\usepackage{booktabs}
\usepackage{lineno}

\title{\boldmath Deep Learning Analysis of Ions Accelerated at Shocks}
\author[a]{P. Swierc,}
\author[a,b]{D. Caprioli,}
\author[c,d]{L. Orusa,}
\author[a]{and M. Cernetic}

\affiliation[a]{Department of Astronomy \& Astrophysics, University of Chicago,\\ 5640 S Ellis Ave, Chicago IL, 60637, USA}
\affiliation[b]{Enrico Fermi Institute, University of Chicago,\\ 5640 S Ellis Ave, Chicago IL, 60637, USA}
\affiliation[c]{Department of Astronomy and Columbia Astrophysics Laboratory, Columbia University,\\ 538 West 120th Street, New York NY, 10027, USA}
\affiliation[d]{Department of Astrophysical Sciences, Princeton University,\\ 4 Ivy Ln, Princeton NJ, 08544, USA}

\emailAdd{pswierc@uchicago.edu}
\emailAdd{caprioli@uchicago.edu}
\emailAdd{luca.orusa@princeton.edu}
\emailAdd{cernetic@uchicago.edu}

\abstract{We study the application of deep learning techniques to the analysis and classification of ions accelerated at collisionless shocks in hybrid (kinetic ions--fluid electrons) simulations.
Ions were classified as thermal, suprathermal, or non-thermal, depending on the energy they achieved and the acceleration regime they fell under. These classifications were used to train deep learning models to predict which particles are injected into the acceleration process with high accuracy ($>90\%$), using only time series of  
the local 
magnetic or electric field they experienced during their initial interaction with the shock. 
An autoencoder architecture was also tested, for which time series of various parameters were reconstructed from encoded representations. 
This study shows the potential of applying  machine learning techniques to extract physical insights from kinetic plasma simulations and sets the groundwork for future applications, including the construction of sub-grid models in fluid approaches.}

\begin{document}
\maketitle
\flushbottom

\section{Introduction}

Diffusive shock acceleration \citep[DSA][]{krymskii1977regular, bell1978acceleration,blandford1978particle, axford1977acceleration} at non-relativistic shocks is a prominent mechanism for producing high-energy particles, hereafter, cosmic rays (CRs).
DSA is a form of Fermi acceleration, in which particles gain energy by being repeatedly scattered across a shock; 
its application to supernova remnants is a promising mechanism to explain the origin of Galactic CRs up to $\lesssim 10^{17}$eV \cite{morlino2012strong,snrcaprioli2012cosmic, hillas05, blasi13}.

DSA's strength is its prediction of a power-law in the momentum-space distribution, which depends only on the shock compression ratio $r = \rho_2/\rho_1$, i.e., the ratio of the downstream to upstream plasma density.
For strong shocks in a monoatomic gas with adiabatic index $\gamma = 5/3$, $r \rightarrow 4$, which yields a momentum slope $q = \frac{3r}{r - 1} \rightarrow 4$.
This generally corresponds to an energy spectrum $\propto E^{-2}$ for relativistic particles.
When the corrections induced by efficient CR acceleration are included \cite{zirakashvili+08, caprioli+08, caprioli+09a, snrcaprioli2012cosmic, haggerty+20, caprioli+20}, spectra become slightly softer ($q\simeq 2.2-3$), in agreement with the phenomenology of shock-powered systems and the CR fluxes  observed at Earth \cite{caprioli15p, diesing+21}.

DSA is a highly non-linear process -- accelerated ions drive upstream turbulence and magnetic field amplification, which further enhances acceleration \cite{reville+12, caprioli2014simulations,caprioli2014simulations2nd}. Therefore, studying the process requires complex modeling that accounts for the non-linear interactions between particles and electromagnetic fields. 
One approach is to simulate a collisionless plasma with fully kinetic codes based on the Particle-In-Cell (PIC) or Vlasov methods \citep{birdsall+91, lipatov02}.
PIC simulations model a plasma using discrete macroparticles that evolve under the Lorentz force on a spatial grid. The particle positions and velocities are used to compute moments of the distribution function, such as charge and current densities, which are deposited onto the grid. These quantities are then used to solve Maxwell’s equations for the electromagnetic fields, which are in turn interpolated back to the particle positions to advance their motion self-consistently.
In contrast, Vlasov methods directly solve the collisionless Boltzmann (Vlasov) equation for the phase-space distribution function of the plasma species, evolving it in time across a multidimensional phase space (typically three spatial and three momentum dimensions, plus time).
Such techniques are able to include all relevant physics \cite[e.g.,][]{birdsall1991particle, bell2006fast, amano2007electron, spitkovsky05, 
sironi+11, niemiec2012nonrelativistic, riquelme2011electron,palmroth2018vlasov}, but require both electrons and ions to be fully resolved. 
Given the order of magnitude difference in proton and electron mass, choosing to resolve the electron scale poses computational limits on the size of the simulation, and importantly the ability to model an ion's long-term evolution.

To solve this, the hybrid approach \cite{lipatov02} treats electrons as a massless neutralizing fluid, while ions are retained as kinetic-macro particles. This approach has successfully been used to model non-relativistic collisionless shocks and DSA \citep[e.g.,][]{winske1985hybrid, giacalone2004large, gargate2007dhybrid, guo2013acceleration, caprioli2014simulations, caprioli2014simulations2nd, caprioli+20, orusa2023fast, orusa2025role,orusa+25a,caprioli+25}. 
Such simulations have revealed that acceleration efficiency at shocks is highly dependent on the inclination of the magnetic field $\theta$ relative to the shock normal. Specifically, DSA only occurs efficiently for parallel to quasi-parallel shocks (i.e., $\theta \lesssim 45^{\circ}$) \cite{caprioli2014simulations2nd, caprioli2014simulations2nd}. Consider instead the case of a perpendicular shock ($\theta = 90^{\circ}$). Here, a particle encountering the shock is confined to gyrate the ordered magnetic field and cannot otherwise enter the upstream. However, during a gyration the particle may still gain momentum $\propto \frac{v}{v_{sh}}$, where $v$ is the initial particle velocity and $v_{sh}$ is the velocity of the shock. This form of acceleration is known as shock drift acceleration (SDA) \cite{ball2001shock}. 
In the case of SDA, the self-generation of upstream turbulence is suppressed, as a particle undergoing SDA will only reach at most one gyro-radius upstream. 
If the shock magnetization is sufficiently low, the direction of the upstream magnetic field is less important and acceleration can proceed beyond a few cycles of SDA and produce power-law ion tails, generally steeper than DSA ones \cite{bell+11, sironi+11}.
In this case, 3D hybrid simulations are needed to capture the cross-field diffusion that allows some protons to come back from downstream \cite{orusa2023fast, orusa2025role,orusa+25a}.  
SDA also plays a crucial role in quasi-parallel shocks, effectively acting as a bootstrap for injection into DSA \cite{caprioli2014simulations2nd, caprioli2014simulations2nd, park+15, gupta+24b}.

Despite significant recent advancements, there are still unanswered questions in shock acceleration theory -- examples include a complete understanding of either electron injection into DSA as a function of shock parameters \citep[see][and references therein]{gupta+25}
or the long-term evolution of quasi-perpendicular and oblique shocks.
In both cases, increasing complexity of simulations poses a bottleneck. For electron injection, this is due to the necessity of resolving the electron scale. For the long-term evolution of quasi-perpendicular and oblique shocks, this is because injection can only occur in a fully 3D simulation \cite{orusa2025role}. 

Additionally, state-of-the-art PIC/hybrid simulations can encompass $>10^{10}$ particles, of which it is reasonable to track and record $>10^6$ at a time. 
However, given this scale, it is not feasible for humans to manually analyze all particle trajectories individually. 
While simulations provide information such as spectra, acceleration efficiency, and magnetic field development very easily, leveraging the well of potentially untapped information requires original avenues, such as machine learning methodology. 

In the last two decades, the use of machine learning in science has experienced significant growth. In the life sciences, for example, some revolutionary applications include protein structure prediction \cite{jumper2021highly, baek2021accurate} or antibacterial discovery \cite{stokes2020deep}. Astrophysics is another one of the prominent fields to adopt such methods \cite{rodriguez2022application}. 
Some of the most common astrophysical subfields in which it is applied include stellar \cite{li2025machine} and galactic analysis \citep[e.g.,][]{gharat2022galaxy, soo2023machine, stein2022mining, khullar2022digs, ciprijanovic2021deepmerge}, as well as cosmology    
\citep[e.g.,][]{dvorkin2022machine, moriwaki2023machine}. 
It has also been used in the context of CRs, with applications such as recreating CR energy from shower data \citep[e.g.,][]{erdmann2018deep,alvarado2023cosmic} or detecting CRs in astronomical images \cite{xu2023cosmic}. Machine learning has also been used in a variety of scientific fields to emulate and speed up otherwise slow simulations \citep[e.g.,][]{kasim2021building,kwan2015cosmic,brockherde2017bypassing}, potentially allowing for results that were previously infeasible due to computational limits. 
Torralba Paz et al. \cite{torralba2025neural} considered a 2D PIC simulation of a perpendicular shock and used a convolutional neural network to predict the maximum energy achieved by electrons during recorded tracks, starting from the corresponding time series of momentum and electric fields. 

In this work, we take the first steps towards applying machine learning to better understand ion injection and acceleration in non-relativistic shocks, considering both quasi-perpendicular and quasi-parallel configurations. 
The paper will be organized as follows -- Section \ref{data_description} describes the two datasets used in the study. Section \ref{methods} describes the three types of deep learning networks employed (two classification networks and one autoencoder network). Section \ref{classification} details the classification experiments on both datasets and Section \ref{ae} describes the autoencoder experiments. Finally, Section \ref{disc} discusses insight into the process of injection into DSA gained through this study, as well as potential applications of machine learning to shock acceleration simulation as a whole. 

\section{Data Set} \label{data_description}

Simulations in this study were conducted using the massively parallel hybrid code {\tt dHybridR} \cite{haggerty2019dhybridr}, the successor of {\tt dHybrid} \cite{gargate2007dhybrid}, which includes the relativistic dynamics of ions, though all the simulations presented here are effectively in the non-relativistic regime.

Shocks are modeled by sending a supersonic flow (along $-\hat{x}$) against a reflecting wall set at $x=0$. Reflected particles then interact with the incoming ones, resulting in the formation of a shock, which propagates away from the wall; transverse boundary conditions are periodic.

For all simulations we retain all the three spatial components of momenta and fields and a set of normalized units are used: distance is measured in units of ion skin depth $d_i\equiv c/\omega_p$, where $c$ is the speed of light and $\omega_p \equiv \sqrt{4 \pi ne^2 / m}$ is the ion plasma frequency, with $n$, $e$, and $m$ referring to the ion number density, charge, and mass respectively;
time is measured in units of inverse cyclotron time $\omega_c^{-1} \equiv mc/eB_0$, where $B_0$ is the magnitude of the initial background magnetic field; 
velocity and energy are normalized respectively to the Alfvén speed $v_A = B_0 / \sqrt{4 \pi mn} = c\omega_c / \omega_p$ and to the kinetic energy of upstream particles $E_{sh} \equiv m(M_Av_A)^2/2$. 
Here $M_A \equiv v_{sh} / v_A$, where $v_{sh}$ is the upstream speed in the simulation frame, is the Alfvénic Mach number.
Since we consider initial Maxwellian ion distributions with thermal speed $v_{th}=v_A$, we also have a sonic Mach number $M_s =  M_A/\sqrt{\gamma}$, where $\gamma=5/3$ is the plasma adiabatic index. 
In the following we will refer to both Mach numbers as $M_A \approx M_s = M$. 
Finally, electrons are assumed to be an adiabatic fluid with $\gamma=5/3$, initially in thermal equilibrium with the ions \citep[][]{caprioli+18}.

Two distinct datasets were generated for this study, corresponding to a perpendicular and a quasi-parallel shock, as discussed below. 

\begin{figure}[tb] 
  \centering
  \includegraphics[scale=.275]{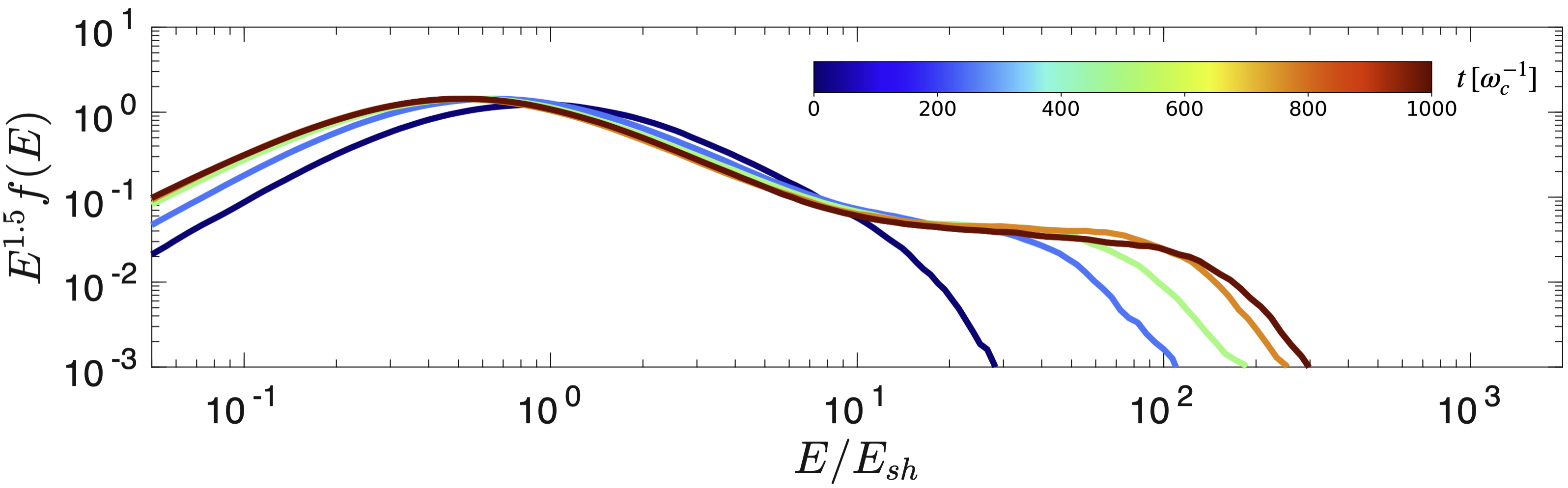}
  \caption{Evolution of the postshock energy spectrum for the 2D parallel $M=10$ shock. The distribution begins as a Maxwellian, but over time it develops a distinct non-thermal power-law tail.}
  \label{fig:spectrum}
\end{figure}

\subsection{Perpendicular Shock Simulation} \label{oblique}
We first consider a perpendicular shock ($\theta = 90^{\circ}$) to perform initial tests with the machine learning methods;
this is a relatively simple regime because only SDA occurs \cite{caprioli2014simulations}. 
The simulation was conducted in 3D  
in a box with size $3000\times 50\times 50 \, (d_i)^3$, with $2.5$ cells per $d_i$, and $8$ particles per cell. 
The much larger size in the $x$ axis is to allow enough travel distance for the shock to develop a sufficient suprathermal population of ions via SDA. The supersonic flow has Mach number $M = 100$. 
The simulation timestep was fixed at $7.5\times10^{-4}\omega_c^{-1}$ and tracks were recorded with a time step resolution of $t = 7.5\times10^{-2}\omega_c^{-1}$. 
The simulation ran for $15 \omega_c^{-1}$, over which about $2.3\times10^5$ particle trajectories were recorded. This limited timescale is due to the 3D simulation setup, which is computationally very expensive. However, for a perpendicular shock it is sufficient to generate a large population of suprathermal particles, since only about one ion gyration is needed, as we will show later.
Following \cite{caprioli2014simulations, caprioli2014simulations2nd}, we define thermal particles as those whose final energy is $E/E_{sh} \leq 2$, about $60\%$ of the recorded ones; 
the remaining $40\%$ encompasses  suprathermal ions with $E/E_{sh} > 2$.

\subsection{Parallel Shock Simulation} \label{parallel}
In the second dataset we simulate the full process of DSA using a parallel shock, with $\theta = 0^{\circ}$ and $M=10$.
In this case, a much larger box is needed to encompass the diffusion length of higher and higher energy ions, and it takes longer for the simulation to develop an extended non-thermal tail, as shown in Figure \ref{fig:spectrum}. 
For these reasons,  the simulation was run in 2D, which is sufficient to capture all the relevant physics  \cite{caprioli2014simulations}. 
The box measures
$40000 \times 250 d_i$, with $2.5$ cells per $d_i$, and $4$ particles per cell. 
The simulation ran for a total of $1000\omega_c^{-1}$, with a timestep of $4.0\times10^{-3} \omega_c^{-1}$. 
We collected $1.2\times10^5$ particle tracks with a time resolution of $8.0\times10^{-2}  \omega_c^{-1}$, with the data evenly split between thermal ($E/E_{sh} \leq 2$), suprathermal ($2 < E/E_{sh} \leq 10$), and non-thermal ($E/E_{sh} > 10$) particles, following the classification in Refs.~\cite{caprioli2014simulations2nd, caprioli2014simulations2nd, johlander+21}. The evolution of the postshock energy spectrum for this simulation is reported in Figure \ref{fig:spectrum}.

\section{Machine Learning Algorithms} \label{methods}
Machine learning is a term that encompasses many computational techniques for data analysis. This study specifically focuses on \textit{deep learning}, a subfield of machine learning in which networks are designed with "hidden" layers between the input and output. 
With the combination of nonlinear functions between layers, deep learning networks are able to use high-level abstraction to learn complex patterns in data. 
This regime makes deep learning ideal for highly complex processes, such as acceleration at shocks. Three types of deep learning models were tested in this study, discussed below. 

\subsection{Convolutional Neural Network}

The model primarily used in this study is a convolutional neural network (CNN) \cite{lecun2002gradient}. CNNs learn filters that extract informative patterns or features from the input data. While they are most commonly employed for image analysis, they can also be applied to other data types such as time series. For more comprehensive overviews, see Refs. \cite{li2021survey, krichen2023convolutional}.

\begin{figure}[!b] 
  \centering
  \includegraphics[scale=.5]{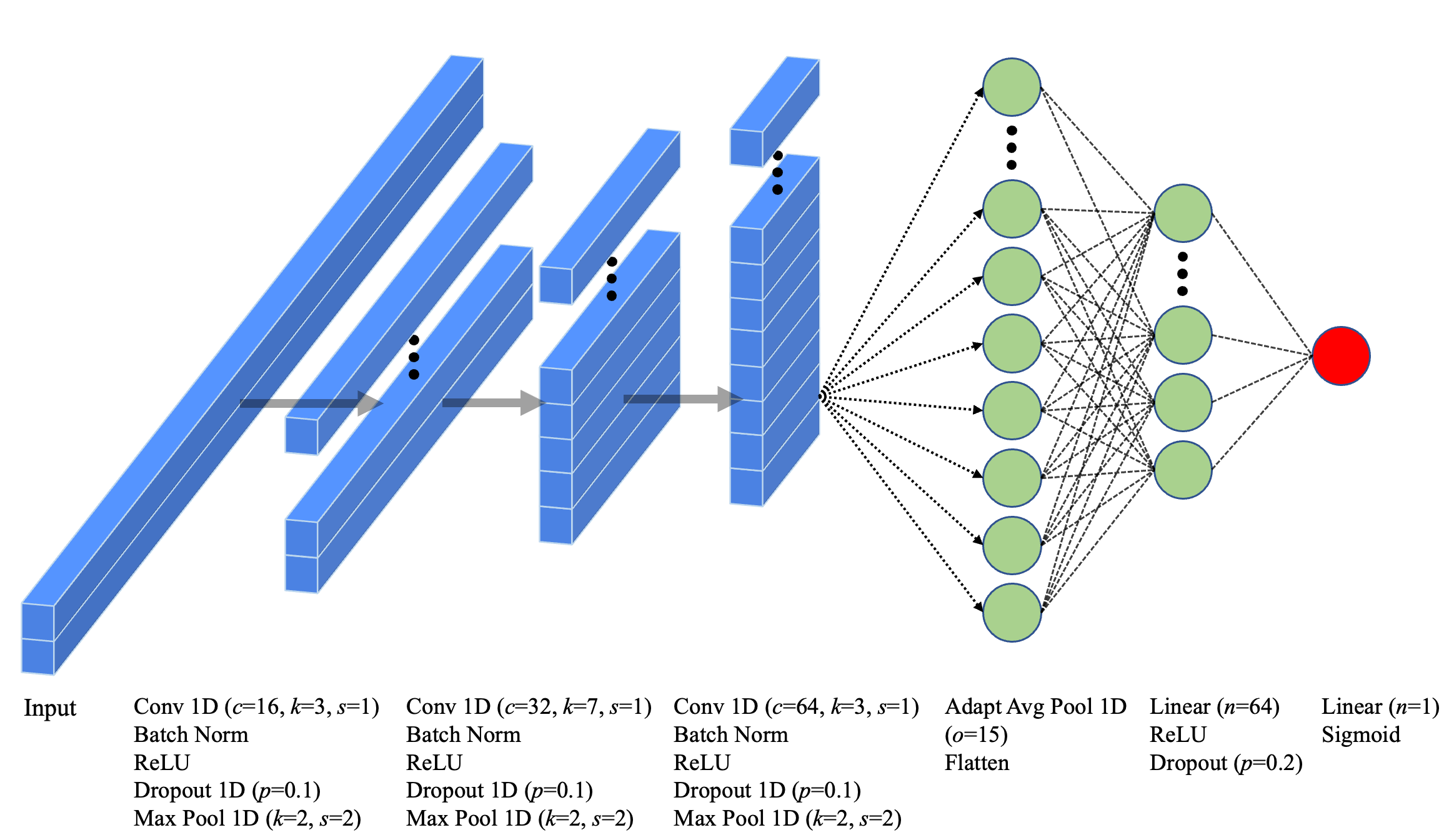}
  \caption{Visualization of the CNN architecture used for classification tasks. Blue blocks represent 1D arrays in convolutional layers, with longer blocks representing arrays of larger lengths. Green circles represent nodes in fully connected layers and the red circle represents the final output of the neural network. Within labels, \textit{c} is the number of output channels from a convolutional layer, \textit{k} is kernel size, \textit{p} is the probability of dropping for a dropout layer, \textit{s} is the stride, \textit{o} is the output size for the adaptive pooling layer, and \textit{n} is the number of nodes in a linear layer. All convolutional layers have a stride of $1$.}
  \label{fig:cnn_vis}
\end{figure}

In this study, we train a CNN to determine whether a particle remains in the thermal population or becomes accelerated. To achieve this, the model takes as input a vector time series corresponding to a particle's initial interaction of the shock. The parameter recorded in this time series is generally the locally experienced magnetic field, though models are trained with other parameters, including the momentum and electric field.
For each input, the network outputs the probability that the particle will undergo acceleration.
The CNN architecture used in this study is visualized in Figure \ref{fig:cnn_vis}.

On a technical level, a CNN operates by extracting features through convolutional layers, which learn kernels (filters) that are convolved with the data in strides. Each convolutional layer is then followed by a block of accompanying layers. Though the ordering of these layers may vary, the sequence we use is batch-normalization $\rightarrow$ activation function $\rightarrow$ pooling. These are defined as follows: the batch-normalization layer helps stabilize training by normalizing the outputs of each layer over a mini-batch of data (the set of data samples used in a training step); an activation function is used to introduce non-linearity, with a common example being the Rectified Linear Unit (ReLU) \cite{nair2010rectified}, which sets negative values to zero; and finally the pooling layer reduces the dimensionality of the data (e.g., by shortening the length of a filtered time series).
A typical CNN architecture includes multiple instances of these blocks, represented by the blue components in Figure \ref{fig:cnn_vis}. After passing through the convolutional blocks, the filtered and reduced data can be flattened into a vector of features. An adaptive pooling layer may also be added before flattening to further compress the feature set.
Following this, fully connected layers (matrix multiplications followed by a non-linear activation) use the extracted features to produce an output; these are represented by the green components of Figure \ref{fig:cnn_vis}. 
For binary classification tasks, a sigmoid function is applied to convert the raw scores (logits) produced by the network into a probability, represented by the final red node in Figure \ref{fig:cnn_vis}. Dropout may also be applied between layers, randomly turning off a fraction of neurons to help prevent the network from overfitting (i.e., memorizing the training data).

To train the network, a loss function must be chosen, which quantifies performance on the given task. Backpropagation, which computes gradients of the loss function with respect to all network parameters, is applied by an optimization algorithm to make updates. For this classification problem, we use the Binary Cross-Entropy (BCE) loss function, which is defined as follows:
for input data with a true label of $y \in \{0, 1\}$, where $0$ corresponds to a thermal particle and $1$ corresponds to an accelerated particle, and an output from the model $\hat{y} = P(y = 1)$, which is the predicted probability that a particle has been accelerated.

We define a particle as accelerated when it hits a threshold energy level, based on \cite{caprioli2014simulations2nd}. This threshold depends on the specific shock (i.e. it is different for the perpendicular and parallel shock datasets described in Section \ref{oblique} and Section \ref{parallel} respectively), and is given when describing the classification experiments for both datasets in Section \ref{sda class} and Section \ref{dsa class} respectively.

Following this, BCE loss is defined as:
\begin{equation}
    \ell_{BCE} = -[y\log (\hat{y}) + (1-y)\log (1-\hat{y})]. 
\end{equation}
Given that CNNs are typically trained in batches of data, the average loss over a batch of $N$ samples is given by:
\begin{equation}
    \mathcal{L}_{BCE} = \frac{1}{N} \sum_{i=1}^{N}\ell_{BCE}(y_i, \hat{y_i}).
\end{equation}
The model is trained by feeding a batch of input data to get a batch of predictions, which are used with the true labels to compute $\mathcal{L}_{BCE}$. The gradient of the loss is then computed and the optimizer is used to adjust model weights as to minimize the loss.

In this study, we only consider the classification problem of whether a particle will be accelerated, rather than the regression problem of predicting the exact energy that a particle will finally achieve \cite{torralba2025neural}. 
This is because the energies of non-thermal particles are always truncated by the finite length of the simulation.
Also, many of the suprathermal particles eventually rejoin the thermal population. 

\subsection{Multilayer Perceptron with Manual Features}

The next network tested in this study, a multilayer perceptron (MLP), is a simpler approach to classification than a CNN. They consist of only fully connected layers (matrix multiplications) with non-linearity functions between them. To this end, they are functionally the same as the final part of a CNN, without any of the preceding convolutional blocks (i.e., just the green nodes in Figure \ref{fig:cnn_vis}). 
MLPs often lack the robustness of CNNs to process higher dimensional and more complex data;
this can be partially remedied by manually choosing features of the input data to give the MLP, instead of directly feeding it more complex data. 
In the context of this study, this means choosing a number of statistics (such as mean, median, or variance) derived from the time series data. These statistics are then fed to the MLP, instead of the entire time series. The same loss function is used for both the CNN and MLP.

\subsection{Autoencoder} \label{ssec:ae}
An autoencoder is a type of neural network that is trained unsupervised, meaning that it does not need access to labeled data. Its general design is to take data (such as particle track time series) and compress it into a certain number of features, then decompress those variables, recreating the original input data. In this way, it learns a lower dimension encoding of more complex data. 
An autoencoder is composed of two parts: an encoder that compresses data into a chosen number of features, and a decoder, which decompresses the features back into a recreation of the inputted data. CNNs are often used as encoders because they are efficient at producing features. For a more comprehensive autoencoder review, see Ref. \cite{berahmand2024autoencoders}.

In this study, we use the CNN architecture in Figure \ref{fig:cnn_vis} for the encoder. The only difference is the final output layer, which consists of 64 features instead of a single node. 
The decoder then takes the same architecture as the encoder, but in reverse, replacing convolutional layers with convolutional transpose layers. Also, pooling layers are removed in the decoder. Instead, the convolutional transpose layers have a stride of 2, meaning that the data is effectively doubled in length after each transpose.

The loss function used for the autoencoder is the mean square error (MSE) loss, which is defined as follows: 
the inputs to this function are the true vector time series, labeled as $\textbf{t}$, and the vector of predictions generated by the model, labeled as $\hat{\textbf{t}}$. 
If $D$ is the total size of $\hat{\textbf{t}}$, then the loss is computed as:
\begin{equation}
    \ell_{MSE} = \frac{1}{D} \sum_{i=1}^{D}(\hat{t}_i - t_i)^2
\end{equation}
and the average loss over a batch of $N$ samples is given by:
\begin{equation}
    \mathcal{L}_{MSE} = \frac{1}{ND} \sum_{n=1}^N \sum_{i=1}^D(\hat{t}_{i,n} - t_{i,n})^2.
\end{equation}

\noindent No normalization is applied to the data during preprocessing.

\subsection{Long Short-Term Memory Network} \label{lstm_back}

Another type of deep learning architecture, known as a Long Short-Term Memory (LSTM) network \cite{hochreiter1997long}, was also considered in this study.
Unlike traditional feed-forward neural networks, such as the aforementioned CNN, where data is processed as a whole, LSTMs take in data sequentially, allowing inference on later data points to be influenced by the context of earlier data points. An LSTM uses internal gates to control what information to keep, discard, and pass on over time, allowing it to maintain relevant context while ignoring irrelevant data.
This usually makes them well fit for time series data, such as particle tracks, though in our specific problem they are not intrinsically better than CNNs for several reasons.

One of the key strengths of an LSTM is its ability to “forget,” or disregard, irrelevant portions of a time series. However, in this study, the data is preprocessed by explicitly identifying the segment corresponding to an ion’s initial interaction with the shock and discarding all preceding information. As a result, the input provided to the model is already restricted to the physically relevant portion of the trajectory. The problem therefore becomes inherently local, with the time series centered on the critical interaction of interest. In this context, the LSTM’s ability to filter out unimportant information is largely unnecessary, as this step is effectively handled during preprocessing based on physical considerations.

Additionally, the time series considered in this study exhibit oscillatory, repeating patterns that persist across the entire window (see, e.g., Fig. \ref{ae_samples}). 
While LSTMs can be applied to such data, CNNs are particularly well suited to identifying recurring local features through their convolutional filters, making them a natural choice for this problem. In contrast, LSTMs would likely offer an advantage for more non-local time series, especially in cases where the data contain extended intervals of limited relevance.

For these reasons, and because CNNs are generally more computationally efficient to train than LSTMs \cite{weytjens2020process, bai2018empirical}, CNNs were chosen as the primary model for this study. Nevertheless, LSTMs achieve comparable performance, as briefly discussed in Section \ref{lstm}.

\section{Classification Experiments} \label{classification}

The classification problem is defined as predicting if an ion will be accelerated based on its initial interaction with the shock. 
This is motivated by the manual inspection of individual ion trajectories in hybrid simulations \cite{caprioli2014simulations2nd, orusa2025role}, which demonstrated that the fate of a particle is determined by its first gyrations around the shock.
Three different types of input data are considered, which include a time series of the ion's momentum ($\mathbf{p}$); the local magnetic field an ion experienced along its path ($\mathbf{B}$); or the local electric field an ion experienced along its path ($\mathbf{E}$). Experiments for both CNN and MLP classification models were conducted on an NVIDIA Quadro RTX 6000 (Turing, 24GB VRAM) GPU. Training one of the CNN models takes approximately $30$ minutes, while training an MLP model takes approximately $15$ minutes.

\subsection{Perpendicular Dataset Classification (SDA)} \label{sda class}

First, we present the experiments conducted with the simpler 3D perpendicular shock dataset, described in Section \ref{oblique}, where the acceleration regime is purely SDA, i.e., particles gain energy by experiencing a net motional electric field while gyrating around the shock  \cite{ball2001shock, park+12, orusa2023fast}. 
For this dataset, we define the two classes as thermal particles ($E/E_{sh} \leq 2$) and  suprathermal particles ($E/E_{sh} > 2$). These energy distinctions are inspired by the fact that in a single cycle of SDA, an ion may approximately double its energy, entering the suprathermal population. 
In all experiments, the time series for $2.3\times10^5$ particles were split via random sampling between training data, validation data used to assess performance during training, and test data reserved for final performance analysis. The split was $80\%$ training, $5\%$ validation, and $15\%$ test.

\begin{figure}[!ht]
   \centering
   \includegraphics[scale=.395]{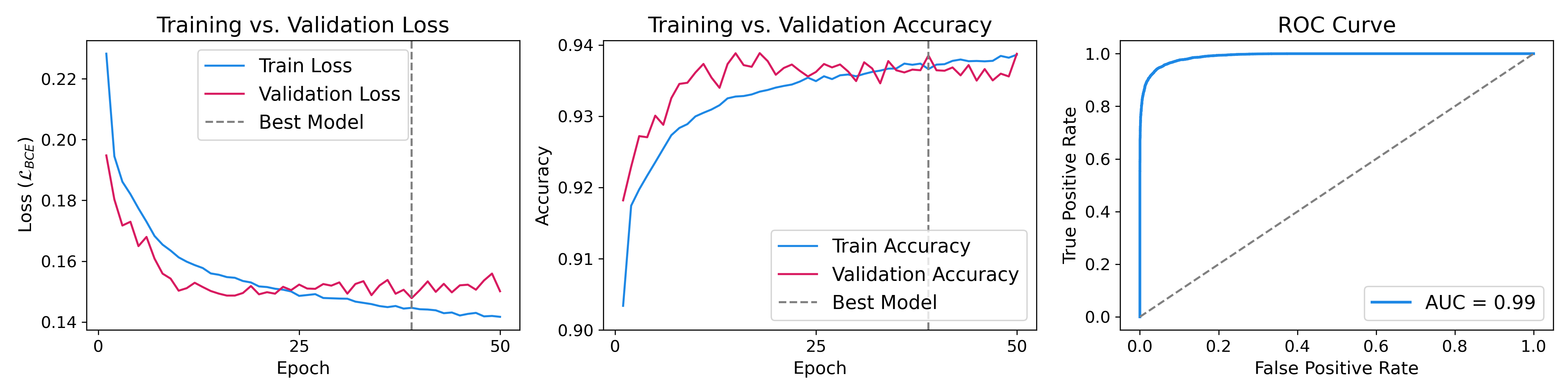}
   \caption{\textbf{Left:} History of training and validation loss for the CNN trained on magnetic field time series from the perpendicular shock (\S\ref{oblique}). Validation loss eventually starts performing worse than training loss, which indicates overfitting. Actual model used in final tests is taken at the best epoch (39). The best model is defined as the model that achieved the minimum validation loss. \textbf{Center:} History of training and validation accuracy. \textbf{Right:} ROC curve for the CNN trained on magnetic field time series from the perpendicular shock.}
   \label{perp_cnn_hist}
\end{figure}

Our first experiments task the CNN architecture reported in Figure \ref{fig:cnn_vis} to predict if a particle will stay in the thermal population or become suprathermal.  
The set of training hyperparameters for the CNN include: a learning rate of $\alpha = 10^{-4}$, which specifies the size of each update to the network's internal parameters; no weight decay, which, when present, pushes model weights towards zero during training to help prevent overfitting; a batch size of $64$, which defines how many samples are processed in one training step; and the Adam optimizer \cite{kingma2014adam}, which dictates how the model updates its internal parameters during training. The optimizer specific hyperparameters were kept at their default values. These include $\beta_1=0.9$, which controls how much the direction of the gradient is smoothed; $\beta_2=0.999$, which controls how much the size of the gradient is smoothed; and $\epsilon=1\times10^{-8}$, which is an added constant to avoid division by zero.
1D dropout layers with a probability of 0.1 are also used within each convolution block and node level dropout layers with a probability of 0.2 are used after the first fully connected layer. The addition of these layers controls overfitting (see Figure \ref{perp_cnn_hist}), where the model begins to memorize training data. The architecture and hyperparameters were chosen through a parameter search on the parallel dataset using Optuna \cite{optuna_2019}, in which the loss on validation data was minimized. The model was optimized with magnetic field data as the input. These choices were also found to be optimal for the perpendicular dataset.

For all types of input data ($\mathbf{p}$, $\mathbf{B}$, and $\mathbf{E}$), the time series were given in 3D. Out of the three types of input data, momentum was used as a baseline test, as it gives the network direct access to how the energy changes over time. The input data of primary focus in this study is then the local magnetic field. Unlike an input of direct momentum time series, one cannot easily derive a particle's energy from solely the magnetic field it experienced, making the classification problem strongly nontrivial. Most diagnostic plots, unless otherwise stated, deal with magnetic field input data. While the electric field input was also tested, it yielded similar results to the magnetic field input in all cases.
For all types of data, the time series consists of the first 86 time steps after the ion encountered the shock, with each step $\Delta t= 7.5\times10^{-2}\omega_c^{-1}$. This corresponds to about the time $2\pi /\omega_c$ it would take for a particle to make one gyration around the shock (i.e., its first interaction). 

The overall performance on test data is reported in Table \ref{perp_stats}. This includes the accuracy as a basic binary accuracy; precision, which is the fraction of positive predictions that were actually positive (i.e., the fraction of predictions that a particle would become suprathermal that were correct); recall, which is the fraction of positive samples that were correctly predicted; and the F$_1$ score, which is the harmonic mean of the precision and recall.

Let \(TP\), \(TN\), \(FP\), and \(FN\) denote true positives, true negatives, false positives, and false negatives. The mathematical definitions are as follows:
\[
\text{Accuracy} \;=\; \frac{TP + TN}{TP + TN + FP + FN},
\qquad
\text{Precision} \;=\; \frac{TP}{TP + FP},
\]
\[
\text{Recall} \;=\; \frac{TP}{TP + FN},
\qquad
F_1 \;=\; 2\cdot\frac{\text{Precision}\cdot\text{Recall}}{\text{Precision} + \text{Recall}}
\;=\; \frac{2TP}{2TP + FP + FN}.
\]

Unsurprisingly, the model trained with momentum time series performs the best, with an accuracy of $98.91\%$. However, the model trained with magnetic field data also performs well, with an accuracy of $94.72\%$. The model trained on electric field data performed comparably to the magnetic field case, with an accuracy of $94.56\%$.

Training and validation accuracy and loss for the model trained with magnetic field data are reported in Figure \ref{perp_cnn_hist}, which also shows the receiver operating characteristic (ROC) curve \ref{perp_cnn_hist}. This metric describes how the model's true positive rate varies with its false positive rate. The closer it gets to forming a right angle at the upper leftmost side of the plot, the better. To achieve this metric, ROC varies the classification probability threshold which is applied to the model outputs. That is to say, it varies the threshold between $[0,1]$ to compute the true positive rate and false positive rate.
The area under the ROC curve (AUC), is another metric shown. The AUC ranges from $0$ to $1$, with a perfect score being $1$.
The ROC curve and AUC confirm that performance was consistent between the two classes, i.e., the model was not significantly better at predicting thermal or suprathermal particles.

\begin{table}[ht]
  \centering
  \begin{tabular}{|l|ccc|ccc|}
    \hline
     & \multicolumn{3}{c|}{\textbf{CNN}} & \multicolumn{3}{c|}{\textbf{MLP}} \\ \hline
                      \textit{Input Data}   & {$\mathbf{p}$} & {$\mathbf{B}$} & {$\mathbf{E}$} & {$\mathbf{p}$} & {$\mathbf{B}$} & {$\mathbf{E}$} \\ \hline
    Accuracy           & $.9891$ & $.9472$ & $.9456$ & $.9308$ & $.8948$ & $.9093$ \\ \hline
    Precision          & $.9892$ & $.9481$ & $.9434$ & $.9262$ & $.8987$ & $.9040$ \\ \hline
    Recall             & $.9882$ & $.9423$ & $.9437$ & $.9318$ & $.8826$ & $.9102$ \\ \hline
    F$_1$-score        & $.9887$ & $.9449$ & $.9436$ & $.9287$ & $.8888$ & $.9066$ \\ \hline
  \end{tabular}
  \caption{\textbf{Perpendicular dataset:} Classification metrics for models trained with time series inputs of either momentum, local magnetic field, or local electric field. CNN models are given on the left and MLP models on the right.}
  \label{perp_stats}
\end{table}

\begin{figure}[]
   \centering
   \includegraphics[scale=.45]{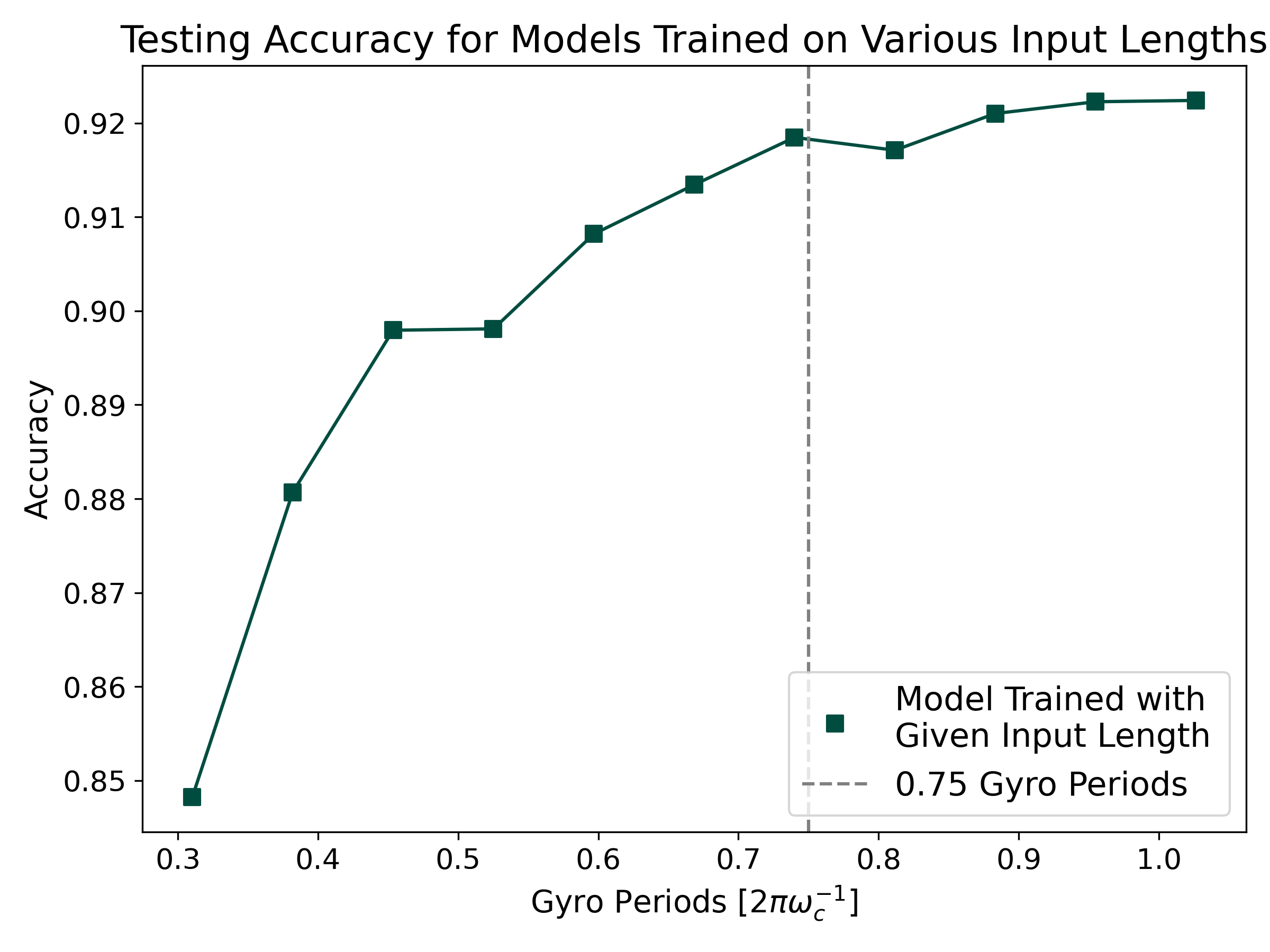}
   \caption{For the \textbf{perpendicular} dataset, with magnetic field input data, this shows multiple CNNs trained with varying number of time steps. Time is given in gyro periods, defined as $2\pi\omega_c^{-1}$. One gyro period is about the time it takes to complete a cycle of SDA. Performance begins to deteriorate at less than $0.75$ gyro periods.}
   \label{perp_cnn_timesteps}
\end{figure}

Finally, we examine how shorter time series inputs affect model performance, in the case of magnetic field input. This is reported in Figure \ref{perp_cnn_timesteps}. Optimal accuracy seems to be reached at around $0.75$ gyro periods. The network maintained good ($>90\%$) accuracy for as few as $0.5$ gyro periods, at which point the accuracy decreased to $85\%$ for as few as 25 time steps. This suggests that the time corresponding to roughly half to three quarters of the first gyration around the shock is most critical for determining if a particle undergoes SDA.

Next, we briefly discuss classification with the MLP network. The problem definition is the same as before but, instead of being fed an entire time series, the MLP is given twelve statistics for each dimension of the time series. These include:
\begin{multicols}{2}
\begin{enumerate}[noitemsep, topsep=0pt]
    \item Minimum value
    \item Maximum value
    \item Mean value
    \item Root mean squared
    \item 25th percentile
    \item 50th percentile
    \item 75th percentile
    \item Mean absolute change
    \item Number of 0 crossings
    \item Number of peaks
    \item Skew
    \item Kurtosis
\end{enumerate}
\end{multicols}

As previously described, the MLP architecture is the same as the fully connected portion of the CNN in Figure \ref{fig:cnn_vis}. It was tested using the momentum, magnetic field, and electric field input option over 86 time steps, for which the aforementioned statistics were calculated in each of the dimensions and fed to the neural network. The same optimizer, batch size, and learning rate as in the CNN were used. The MLP was able to achieve an accuracy of $89.48\%$ on the test set (Table \ref{perp_stats}) for magnetic field input data, meaning it underperformed compared to the CNN. However, this is still a relatively high accuracy given the significant simplification in network complexity compared to the CNN.

\subsection{Parallel Dataset Classification (DSA)} \label{dsa class}

Next, we present the classification experiments conducted with the more complex 2D parallel shock dataset, described in Section \ref{parallel}. 
For this dataset, we define two classes: thermal+suprathermal particles ($E/E_{sh} \leq 10$) and non-thermal particles ($E/E_{sh} > 10$), i.e., the particles that join the DSA tail, which grows with time \cite{caprioli2014simulations2nd}. Suprathermal particles are considered together with thermal particles since, given enough time, most of them 
are thermalized, rather than being injected into DSA.
For typical quasi-parallel shocks, about $25\%$ of the ions crossing the shock become suprathermal, compared to $\sim 1\%$ that achieve non-thermal energies \cite{caprioli2014simulations2nd}.

For classification with a CNN network, the same model architecture, hyperparameters, and loss function were used as in the perpendicular case: input time series were used to cast the binary prediction of whether a particle reached a non-thermal energy. Performance based on the three types of input time series ($\mathbf{p}$, $\mathbf{B}$, and $\mathbf{E}$) are given in Table \ref{par_stats}. The model with momentum inputs performed the best with an accuracy of $92.56\%$, but the model with magnetic field inputs was also able to achieve an accuracy of $91.70\%$. The model with electric field inputs performed slightly better, with an accuracy of $92.24\%$.

\begin{figure}[]
   \centering
   \includegraphics[scale=.45]{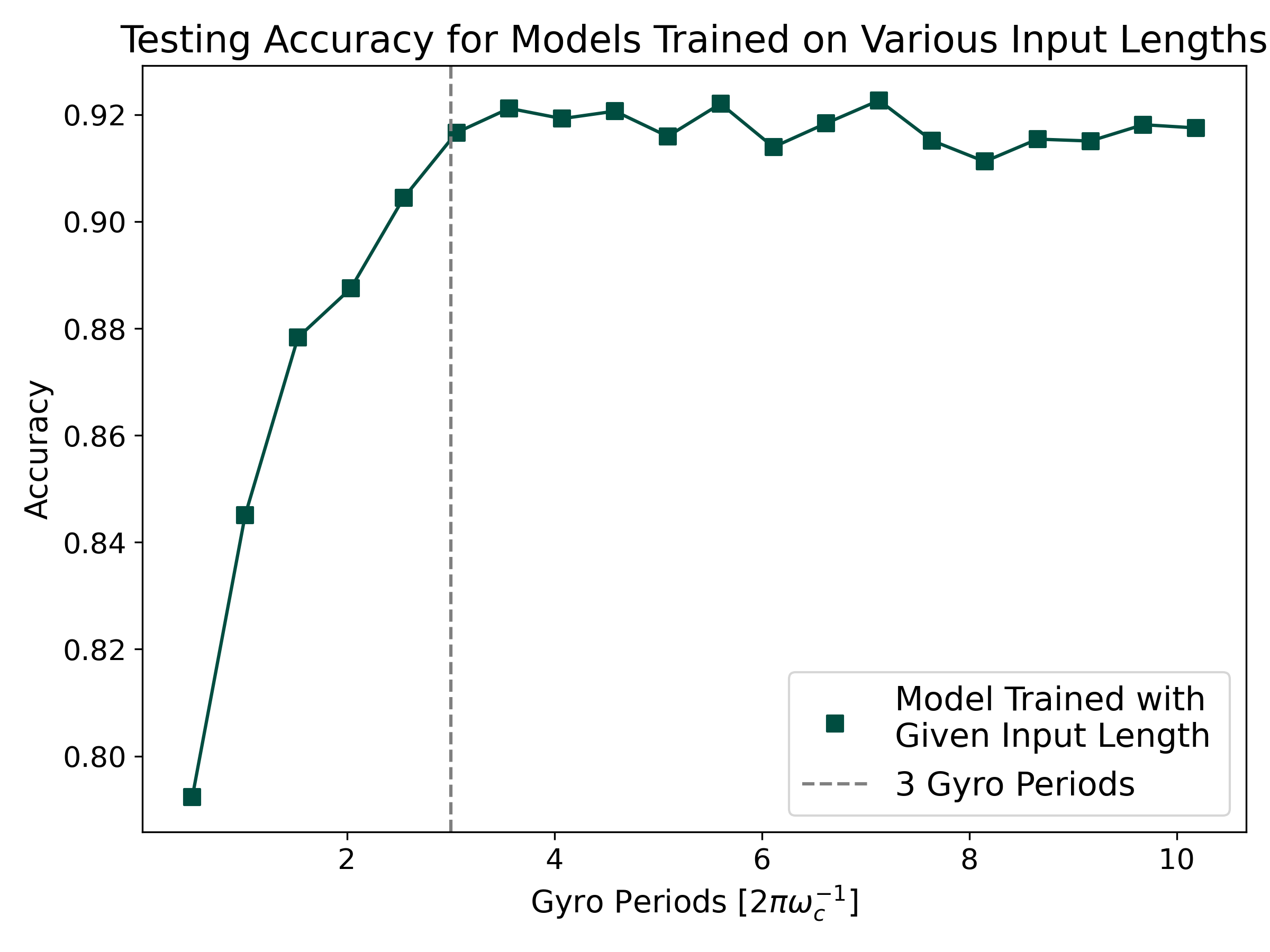}
   \caption{
   Accuracy of multiple CNNs trained with varying numbers of magnetic field time steps for the parallel dataset. Performance severely deteriorates if less than three gyro periods are used. }
   \label{par_cnn_timesteps}
\end{figure}

For all three CNN models, longer time series of 238 time steps were used in training, which roughly corresponds to three gyro periods. Figure \ref{par_cnn_timesteps}, which reports the testing accuracy of models trained on time series of varying lengths, demonstrates that this is a threshold for good accuracy ($>90\%$). The potentially interesting physical implications of this threshold are discussed in Section \ref{inj}. It should also be noted that the majority of non-thermal particles (approximately $55\%$) does not reach $E/E_{sh} \geq 10$ within the 238 time steps given. This means that it is a true predictive problem for those particles, even with momentum time series inputs.

\begin{figure}[]
   \centering
   \includegraphics[scale=.395]{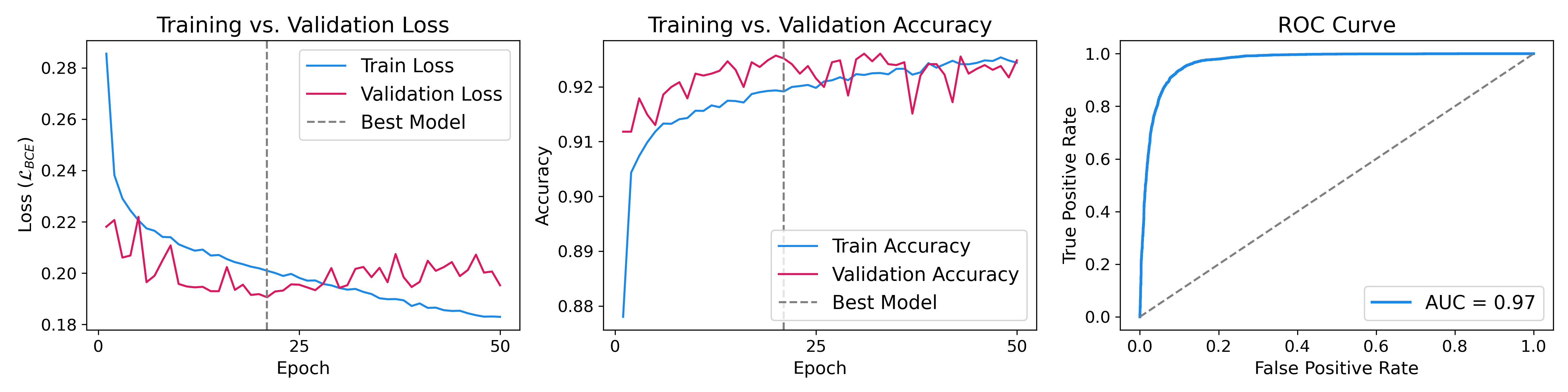}
   \caption{\textbf{Left:} History of training and validation loss for the CNN trained on magnetic field time series from the parallel dataset (\S\ref{parallel}). Actual model used in final tests is taken at the best epoch (21). \textbf{Center:} History of training and validation accuracy. \textbf{Right:} ROC curve, with AUC shown in legend.}
   \label{par_cnn_hist}
\end{figure}

The training history (Figure \ref{par_cnn_hist}) and ROC curve (Figure \ref{par_cnn_hist}) are also reported for the CNN model trained on magnetic field inputs. This model suffers from more severe overfitting compared to the perpendicular case, likely due to the smaller dataset ($1.2\times10^5$ particles with the same training, validation, and test split as in the perpendicular dataset, that had $2.3\times10^5$ particles).

\begin{table}[ht]
  \centering
  \begin{tabular}{|l|ccc|ccc|}
    \hline
     & \multicolumn{3}{c|}{\textbf{CNN}} & \multicolumn{3}{c|}{\textbf{MLP}} \\ \hline
                      \textit{Input Data}   & {$\mathbf{p}$} & {$\mathbf{B}$} & {$\mathbf{E}$} & {$\mathbf{p}$} & {$\mathbf{B}$} & {$\mathbf{E}$} \\ \hline
    Accuracy             & $.9256$  & $.9170$  & $.9224$  & $.9227$ & $.8986$ & $.8665$ \\ \hline
    Precision            & $.9168$  & $.9029$  & $.9135$  & $.9152$ & $.8867$ & $.8539$ \\ \hline
    Recall               & $.9055$  & $.9101$  & $.9058$  & $.9079$ & $.8800$ & $.8354$ \\ \hline
    F$_1$-score          & $.9108$  & $.9064$  & $.9095$  & $.9114$ & $.8832$ & $.8435$ \\ \hline
  \end{tabular}
  \caption{\textbf{Parallel dataset:} Classification metrics for models trained with time series inputs of either $\mathbf{p}$, $\mathbf{B}$, or $\mathbf{E}$. CNN models are on the left and MLP on the right.}
  \label{par_stats}
\end{table}

The same MLP network as in Section \ref{sda class} was also tested on the parallel dataset. For the two dimensions of either $\mathbf{p}$, $\mathbf{B}$, or $\mathbf{E}$, the same statistics as in the previous section were used as features, calculated over $238$ time steps. 
The MLP network performed with slightly reduced accuracy compared to the CNN; 
with the magnetic field time series input, it achieved a testing accuracy of $89.86\%$ (Table \ref{par_stats}). 
Compared to the CNN and MLP networks trained on the perpendicular dataset, those trained on the parallel dataset have a smaller gap in performance. 
This discrepancy arises because the MLP trained on the parallel or perpendicular datasets performs similarly,
whereas the CNN in the parallel case underperforms compared to its perpendicular counterpart.
This could be due to the CNN overfitting concern: the CNN trained on the parallel dataset overfits more severely compared to the one trained on the perpendicular dataset, leading to worse performance. For the MLP networks, however, overfitting is less of a concern with either dataset, due to the smaller and more simple network architecture, as well as the reduced size of the input data (12 summary statistics instead of the entire time series). 
Additionally, the more complex nature of the problem considered for the parallel shock dataset likely makes it more difficult to classify the final population of most challenging tracks, contributing further to the lower CNN accuracy on the parallel dataset.

\subsection{Classification with Long Short-Term Memory} \label{lstm}

As discussed in Section \ref{lstm_back}, CNNs were chosen as the primary model of focus for classification in this study, instead of LSTMs. Performance between these two architectures, though, was found to be highly comparable. The testing accuracy for LSTM models on the classification problems described in Sections \ref{sda class} and \ref{dsa class} are both given in Table \ref{lstm_stats}. For all models, the CNN and LSTM performed within a $1\%$ difference in accuracy (compared against Tables \ref{perp_stats} and \ref{par_stats}). 
It is then likely that either architecture is sufficient to converge to a similar solution for the problems in this study, though we note that the maximum accuracy achievable may not be intrinsic in the problem but rather a reflection of the level of refinement and the size of the training set that the modeler can afford.

\begin{table}[]
  \centering
  \begin{tabular}{|l|ccc|}
    \hline

                      \textit{Input Data}   & {$\mathbf{p}$} & {$\mathbf{B}$} & {$\mathbf{E}$} \\ \hline
    Accuracy on Perpendicular Dataset            & $.9857$ & $.9410$ & $.9371$  \\ \hline
    Accuracy on Parallel Dataset           & $.9180$ & $.9186$ & $.9249$  \\ \hline
  \end{tabular}
  \caption{Classification accuracy for LSTM models trained with various time series inputs of either momentum, local magnetic field, or local electric field. Performance on the perpendicular dataset problem described in Section \ref{sda class} and parallel dataset problem described in Section \ref{dsa class} are both given.}
  \label{lstm_stats}
\end{table}

\section{Autoencoder Experiments} \label{ae}

We considered the autoencoder architecture discussed in Section \ref{ssec:ae}, using as input time series of $\mathbf{p}$, $\mathbf{B}$, and $\mathbf{E}$;
for each of them, we tried to recreate all of the three time series as output. For each combination, a separate model was trained.

\begin{figure}[b]
   \hspace*{-0.5cm}
   \centering
   \includegraphics[scale=.45]{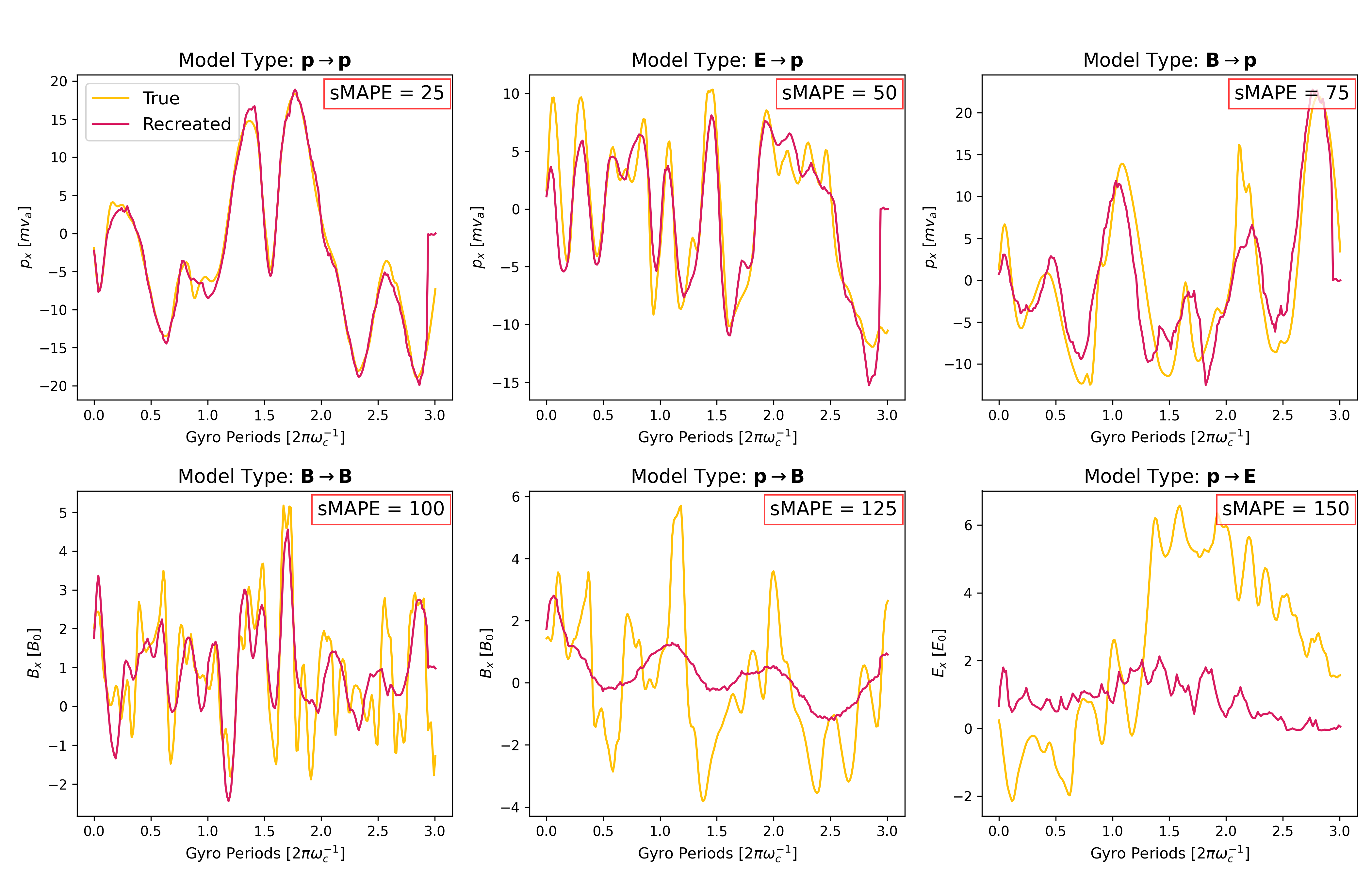}
   \caption{Reference samples from the parallel case, demonstrating performance at various sMAPE values in the $x$ dimension. The model type in the titles describes input $\to$ output parameter. 
   Sample recreations with low sMAPE values are able to capture morphology, but at the highest values, little to no morphology is captured.
   For these samples, the sMAPE was only computed via the model output in the $x$ dimension, as to match the visualized plots.}
   \label{ae_samples}
\end{figure}

Hyperparameters were kept the same as with classification, save for the loss function (MSE) and learning rate, which was increased to $\alpha = 3\times 10^{-3}$. 
Dropout was only used in the encoder and not the decoder. 
A smaller hyperparameter optimization over validation loss for these choices was again run with Optuna \cite{optuna_2019}, but only on the baseline model trained to take an input of $\mathbf{p}$ and recreate that same time series on the parallel dataset. 
For the perpendicular case, all models were tasked with recreating time series in 3D over the first $86$ time steps after an ion encountered the shock. This length was chosen because it corresponds to the time necessary to optimally predict if a particle is accelerated via SDA, as discussed in Section \ref{sda class}. 
The autoencoder experiments were repeated for the parallel dataset with everything kept the same, save for the time series being in 2D and of length 238 (\S\ref{dsa class}). All autoencoder experiments were conducted on an NVIDIA Quadro RTX 6000 (Turing, 24GB VRAM) GPU. Training one of the autoencoder models takes approximately $45$ minutes.

Because MSE is a scaled value and differs greatly based on the parameter that is being recreated, the metric we use to compare autoencoders is the symmetric mean absolute percentage error (sMAPE). 
This gives a percentage or relative error, scaled from $0\%$ to $200\%$, with a lower score corresponding to better performance (i.e., $0\%$ is a perfect score). 
Over $N$ samples, the sMAPE is given by
\begin{equation}
    \mathcal{L}_{\mathrm{sMAPE}}
\;=\;
\frac{100\%}{N D}
\sum_{n=1}^{N}\sum_{i=1}^{D}
\frac{2\big|\hat{t}_{i,n} - t_{i,n}\big|}
     {\big|\hat{t}_{i,n}\big| + \big|t_{i,n}\big|}
\end{equation}

The symmetric version was chosen over the regular mean absolute percentage error because the latter blows up when the true time series approaches zero, even for models with good performance. All of the data used in this study oscillates around zero, so this happens often. Samples of recreated time series that scored various sMAPE values are shown in Figure \ref{ae_samples}. 

\begin{figure}[]
   \hspace*{-1.2cm}   
   \centering
   \includegraphics[scale=.398]{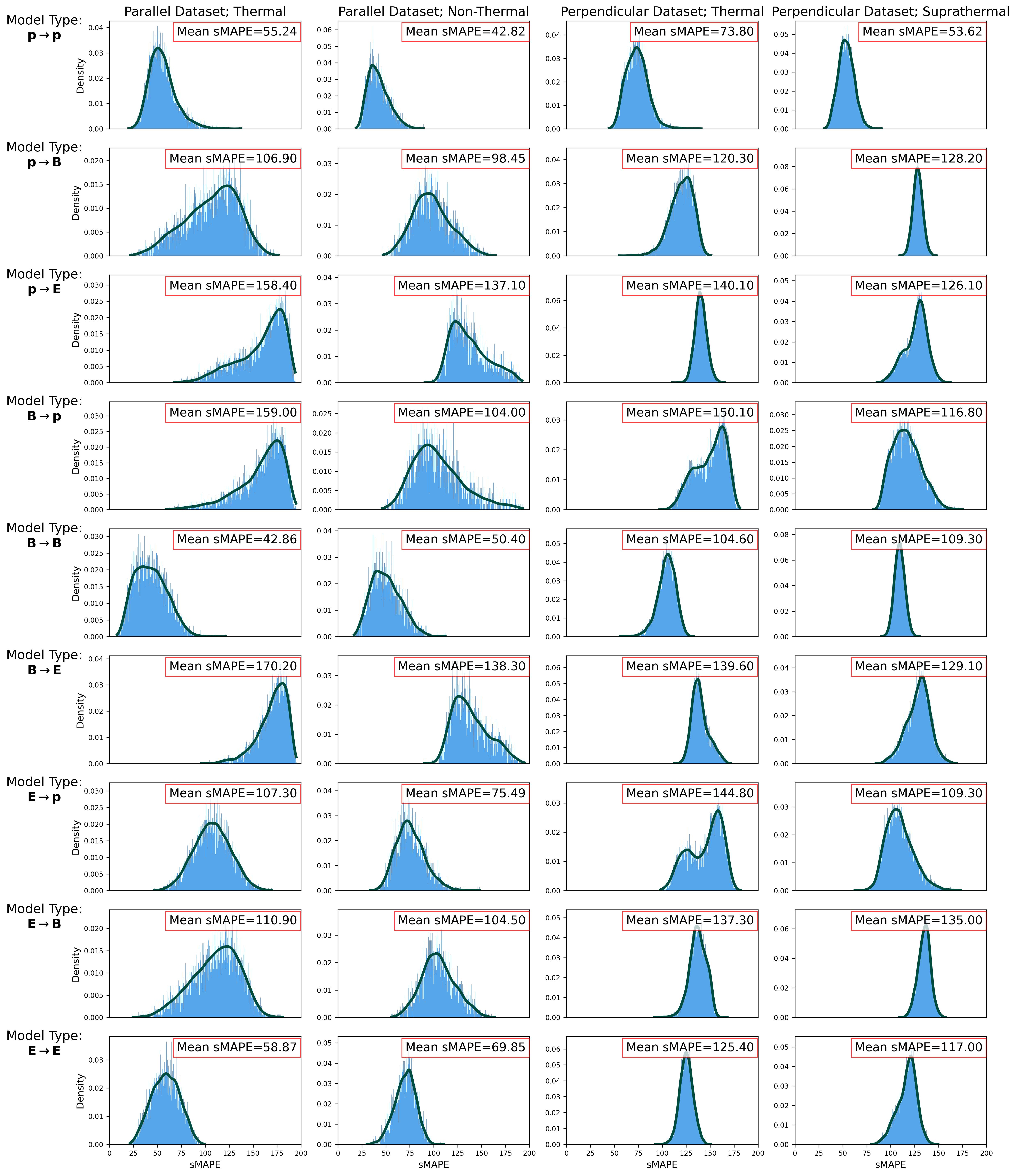}
   \caption{Test dataset distribution of sMAPE scores for all networks. Distributions are split between the thermal and accelerated particle populations of both datasets, with each column representing one such population. Each row is a distinct model type, defined by the type of input and output data the model was tasked with recreating.
   A kernel density estimation (KDE) of the distribution is plotted over the binned density values.}
   \label{ae_grid}
\end{figure}

The sMAPE performance of all nine autoencoders for both parallel and perpendicular cases is reported in Figure \ref{ae_grid}.
Each panel shows the distribution of scores each autoencoder, and the mean sMAPE over the testing dataset is given in the legend; 
performance across the thermal and accelerated particle populations is also separated, to check possible biases. 

In general, the majority of models perform better when recreating the tracks of accelerated particles, which is the focus of the study. 
Strong performance on the parallel dataset is also preferable over the perpendicular dataset, because acceleration via DSA is of higher interest than SDA. This is why the autoencoder setup was optimized on the parallel dataset. 

\begin{figure}[b]
   \centering
   \includegraphics[scale=.5]{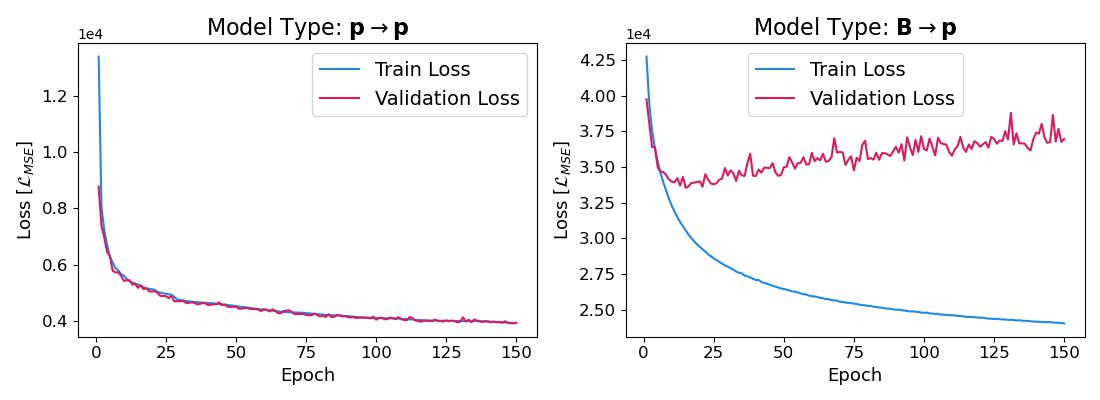}
   \caption{\textbf{Left:} History of training and validation loss for the autoencoder trained to recreate momentum time series from inputs of the same momentum time series, with data from the parallel shock dataset (\S\ref{parallel}). Validation loss decreases with training loss, indicating little to no overfitting. This model performed better than other models, with a mean sMAPE of 42.82 for non-thermal particles in the test set (Figure \ref{ae_grid}). \textbf{Right:} Loss history for the autoencoder trained to also recreate momentum time series in the parallel dataset, but from inputs of magnetic field time series. Validation loss quickly starts performing far worse than training loss, which indicates severe overfitting. This model performed worse than model shown on the left, with a mean sMAPE of 104.00 for non-thermal particles in the test dataset (Figure \ref{ae_grid}).}
   \label{ae_hist}
\end{figure}

For the parallel case, models that take and recreate the same type of data (either $\mathbf{p}$, $\mathbf{B}$, or $\mathbf{E}$) perform well across the board, with mean sMAPE scores $<75$. For a qualitative visualization of this score, see Figure \ref{ae_samples}.
Additionally, the model that takes electric field input data and recreates momentum time series performs well on the non-thermal population of particles in the parallel dataset, with a score of $75.49$. 
The models that take momentum and recreate magnetic field time series; take magnetic field and recreate momentum time series; and take electric field and recreate magnetic field time series are all able to capture some general morphology as well, with scores around $\approx 100$ on the non-thermal population of the parallel dataset (see Figure \ref{ae_samples} for a qualitative visualization of this score). 
For the perpendicular dataset, the only strong performance was the baseline model that both took and recreated momentum time series. However, it should be noted that the autoencoder architecture was only optimized for the base case of momentum input data being used to recreate the same momentum time series for the parallel dataset. The additional results on other problem setups are shown, but these are likely suboptimal results. Overfitting posed a significant bottleneck on the poorer performing models (Figure \ref{ae_hist}), so future studies that work with larger datasets and optimize architectures for each type of data may yield better results. Nonetheless, it is significant that even the unoptimized models with electric field and magnetic input data were able to recover morphology of momentum time series they never had access to, for particles accelerated via DSA in the parallel dataset.

\section{Discussion} \label{disc}

\subsection{Ion Injection at Shocks} \label{inj}

An important result of this study is the insight it gives into the process of injection into DSA. 
Ref.~\cite{caprioli2014simulations2nd} puts forward a minimal model that explains ion injection for quasi-parallel shocks, which was based on the analysis of many individual trajectories in a hybrid simulation.

This study proposed that the first two/three gyrations after the first encounter with the shock are crucial to determine the fate of a particle in DSA. We locate a particle's first encounter with the shock by determining the first point in time at which $B \geq 5 B_0$, where $B$ is the modulus of the local magnetic field at that point in time and $B_0$ of the initial background magnetic field.
After an initial reflection off the quasi-periodically reforming shock barrier \cite{thomas+90}, which occurs for about $25\%$ of the particles impinging on the shock, particles gain energy and become suprathermal. 
Depending on the shock inclination, some of them can outrun the shock and escape upstream, while the rest is re-caught by the shock. 
Every shock encounter is lossy, in the sense that a smaller and smaller fraction of the ions is able to perform more and more gyrations.
Above some critical energy $\bar{E}(\theta)$ all particles can overrun the shock and must rely on diffusion upstream to go back, effectively entering DSA. 
For all quasi-parallel shocks $\bar{E}\sim 10 E_{sh}$, which justifies our choice for the non-thermal threshold.

This theory is reflected perfectly in the classification results with the two datasets. For the perpendicular shock, in which particles are only accelerated to suprathermal status via SDA, strong performance is achieved with as little as half a gyro period of time series data, corresponding to a single interaction with the shock. 
On the other hand, at parallel shocks at least three gyro periods of time series data are required to get comparable classification results. This corroborates the findings that an initial 2–3 cycles of SDA are critical for a particle to be injected into the full DSA process and that the initial reflection alone is not sufficient for injection.

While the physical implications of the proof-of-principle analysis presented here offer a new, model-agnostic, take on the already-solved problem of ion injection, the technique is promising for unraveling much more complicated problems, such as electron injection.
In fact, despite the progress that came from full-PIC simulations, \cite[e.g.,][]{amano2007electron, sironi+11, guo+14a, guo+14b, park+15, xu+20, bohdan+19a, marcowith+20, gupta+24b, gupta+25}, a comprehensive theory for the relative normalization of electron and ion power-law tails is still missing.
It also remains to be assessed whether there are instances where electron DSA alone may occur \cite{brunetti+14}.

\subsection{Construction of Kinetic Sub-Grid Models}

Using autoencoders, we found that the time series for some key parameters defining the track of a particle undergoing acceleration may be compressed and recreated. A possible alternative would be transitioning to a variational autoencoder (VAE) \cite{kingma2013auto}. This architecture replaces the discrete feature encoding with a continuous distribution, which may be sampled to generate new, synthetic data. 
However, in this study, the autoencoders struggled to recreate some of the parameters with more complex time series. Overfitting was an issue in training such autoencoders, meaning that larger datasets may be able to improve performance. This, along with the use of VAEs, will be the topic of future study.

A potentially interesting application of machine learning to DSA simulations would be to add sub-grid kinetic physics to fluid treatments, e.g., the electron fluid in hybrid, or the thermal plasma in MHD-PIC approaches, in which a kinetic CRs population is superimposed on a magneto-hydrodynamical fluid \citep[e.g.,][]{lucek2000non, bai2015magnetohydrodynamic,mignone+18, van2018magnetic,dubois2019shock, sun+23}. 
These kinds of codes can follow the evolution of a collisionless shock for more macroscopic timescales than PIC codes, but requires a prescription for the injection of macroparticles at the shock, a non-trivial task especially for high-Mach cases in which the shock transition is spread over many cells.
Machine learning tuned to kinetic simulations, possibly with coarse sampling to mimic the reduced MHD-PIC resolution and/or MLP architectures, looks like a very promising way to handle particle injection in an agnostic but self-consistent way; 
a similar approach could be used to include electron injection and acceleration into hybrid simulations.
These will be the object of future studies.

\section{Conclusions}

This study puts forward the first  ---to our knowledge--- attempt to analyze injection and acceleration at non-relativistic collisionless shocks with deep learning methods. 

We constructed a convolutional neural network (CNN) able to predict with high accuracy ($>90\%$) whether a particle from a simulated shock dataset is going to be accelerated out of the thermal population, utilizing only a time series of the locally experienced magnetic or electric field. 
For perpendicular shocks, a time series of about one gyro-period that starts at the first interaction of the particle with the shock can classify thermal and suprathermal particles with an accuracy $>94\%$ (\S\ref{sda class}).
For quasi-parallel shocks, a longer time series of about 3 cyclotron times is needed to achieve an accuracy $>90\%$ in classifying thermal and non-thermal particles, namely those who are injected into DSA. 
These findings confirm in an independent way the results obtained with manual inspection of particle trajectories that 
 the first 2-3 interactions with the shock are critical for a particle to be promoted into DSA \cite{caprioli2014simulations2nd}.

We have also experimented with a multilayer perceptron (MLP) network fed with reduced statistical properties of the time series, rather than the time series themselves. 
While generally performing worse than our CNN, this neural network can still achieve $\sim 85-93\%$ accuracy in the cases considered, which suggests that ---with further optimizations--- it may be possible to predict particle injection from coarser samples of the shock electromagnetic structure.

Additionally, an autoencoder's ability to capture the  morphology of momentum, magnetic field, and electric field time series of particles accelerated via DSA was demonstrated. 
Each autoencoder model was trained to take one type of data (momentum, electric field, or magnetic field time series) as input and another as output.
Some of these configurations performed much better than others, with the errors quantified in Figure \ref{ae_grid}.

These results demonstrate the potential success of machine learning applied to understanding particle acceleration at shocks. 
As computational techniques and simulations further develop, we expect that deep learning techniques will facilitate the continued extraction of key insight into the process of acceleration at shocks and the development of subgrid models for embedding kinetic physics into fluid plasma simulations.

\acknowledgments
The authors thank Aleksandra Ciprijanovic and Harley Katz for their advice and suggestions.
They also thank the referee for their helpful comments, as well as  
the University of Chicago Research Computing Center for providing the computational resources to conduct this research.
P.S. was supported by the University of Chicago Jeff Metcalf Internship Program.
D.C.~was partially supported by NASA (grants 80NSSC24K0173 and 80NSSC23K1481) and NSF (grants AST-2510951 and AST-2308021). L.O. acknowledges the support of the Multimessenger Plasma Physics Center (MPPC), NSF grants PHY2206607 and PHY2206609.

\bibliographystyle{JHEP}
\bibliography{Total.bib, main}

\end{document}